\documentclass{aa} 
\usepackage{graphicx}
\begin{document}
   \title{
A deconvolution-based algorithm for crowded field photometry with
unknown Point Spread Function}

   \titlerunning{Crowded fields Point Spread Function and deconvolution}

   \author{P. Magain\inst{1}  
           \and F. Courbin\inst{2}  
           \and M. Gillon\inst{3,1}
           \and S. Sohy\inst{1}  
           \and G. Letawe\inst{1}
           \and V. Chantry\inst{1}
           \and Y. Letawe\inst{1} 
          }

\offprints{P. Magain (Pierre.Magain@ulg.ac.be)}

\institute{Institut  d'Astrophysique et de G\'eophysique, Universit\'e  de
  Li\`ege, all\'ee du 6 Ao\^ut 17, B-4000 Li\`ege, Belgium 
  \and
 Ecole Polytechnique F\'ed\'erale de Lausanne (EPFL),
 Laboratoire d'Astrophysique,
 Observatoire,
 CH-1290 Sauverny,
 Switzerland
  \and
 Observatoire de Gen\`eve,
 51 Chemin des Maillettes,
 CH-1290 Sauverny,
 Switzerland
}

   \date{Received ; accepted }

   \abstract{ A  new method is   presented for determining  the  Point
     Spread Function (PSF) of images  that lack bright and isolated
     stars.   It is based  on the same  principles as the MCS (Magain,
     Courbin, Sohy, 1998) image deconvolution algorithm. It uses the
     information contained in all stellar images to achieve
     the double task of  reconstructing the PSFs for single or multiple
     exposures of the same  field and to extract the photometry of  
     all point sources  in  the field of view.  The use of the full
     information available allows to construct an accurate PSF.  The
     possibility to simultaneously consider several exposures makes it
      very  well suited  to  the measurement of the
     light curves   of blended point  sources from  data that would be
     very difficult or even impossible to analyse with traditional PSF
     fitting  techniques.  
     The  potential  of  the method  for the
     analysis of ground-based    and space-based data is  tested on
     artificial images and illustrated by several examples, including
     HST/NICMOS images of a lensed quasar and VLT/ISAAC images of
     a faint blended  Mira star in  the  halo of the  giant elliptical
     galaxy   NGC~5128   (Cen   A).
      }

   \maketitle
%

\section{Introduction}

In recent years, the importance of point sources CCD photometry has
continuously grown.  Among other applications, let us mention the
determination of colour-magnitude diagrams for stellar clusters, the
study of variable stars in clusters or in external galaxies and the 
search for microlensing events and for planetary transits in light curves, by 
experiments such as OGLE (e.g. Udalski et al.\cite{Udalski}).

When the stars are sufficiently isolated, aperture photometry may 
provide quite reliable results.  However, this is seldom the
case, and techniques based on Point Spread Function (PSF) fitting have to
be used when the stellar images overlap.  The most popular of these
crowded field photometry methods is undoubtedly DAOPHOT (Stetson
\cite{Stetson}).  In such methods, the PSF is generally determined
from the shape of sufficiently isolated point sources.

However, it may happen that the crowding of the field is such that
no star is sufficiently isolated to allow a reliable PSF determination.
This may severely affect the photometric precision as any error in
the PSF will translate in photometric errors for blended stars, which
will be accurately separated only if the PSF used in the fitting procedure
is the correct one.

In such very crowded fields, it appears therefore necessary to use
a PSF determination method which is able to give accurate results
even in situations where all stars are blended to some degree.  This
is the aim of the present paper, which combines image deconvolution
and PSF determination for reaching better performances in both areas
and, as a consequence, for providing excellent photometric accuracy.

Indeed, it has also recently become clear that astronomical  image
deconvolution  is a   useful tool for   reaching  higher spatial
resolution  and extracting  quantitative  and precise information from
the data.  It should not simply be considered as a  way to improve low
quality images, or as a competitor to other techniques but rather as a
complementary method,  which better allows  to take full  advantage of
the existing or future observational techniques.

For example,  the  Hubble Space  Telescope  (HST),  which has a  Point
Spread Function (PSF) with a very compact core, still suffers from low
frequency blurring due to its extended and complex wings. This is also
the case for PSFs obtained with  adaptive optics intruments and will
be true with the   future Extremely Large Telescopes, all  using
adaptive optics. These   extended wings (and  in  fact  also  the high
frequency signal due to the core of the PSF)  can often be efficiently
removed  by  adequate  deconvolution methods,   at    a cost  which  is
negligible compared  to the investment  in the instrument itself.

Similarly to crowded field photometry, the quality of the deconvolution
critically depends on the accuracy of the PSF determination.  The present
paper describes a method which simultaneously allows to obtain accurate PSFs
and to solve the deconvolution problem in fields dominated by point sources,
even when the crowding is so severe that all stars are significantly blended.

\section{Mathematical context}

While  deconvolution   can be  a  powerful technique,    it  is also a
mathematically ill-posed problem.   Many algorithms have been proposed
to deconvolve images, but generally with rather  modest success.  In a
previous  paper (Magain, Courbin and  Sohy \cite{MCS}; hereafter MCS),
we have shown that one of the main  problems with the existing methods
is that they try to recover the light distribution at full resolution,
i.e.\ as it would be obtained with  a perfect instrument (e.g. a space
telescope   of infinite size).   As   the  light distribution is   not
modelled as a continuous function,   but represented on a pixel  grid,
with finite  pixel size $\Delta x$,  the sampling theorem implies that
only  frequency  components up   to  the Nyquist  frequency  $(2\Delta
x)^{-1}$  can be correctly taken  into  account.  Components of higher
frequency  are mixed up with the  lower frequency ones by the aliasing
phenomenon and are responsible for some of  the artefacts which appear
when using most deconvolution techniques.

In MCS, we have shown that better results can be  obtained if one does
not attempt to recover the light  distribution at infinite resolution,
but  rather at an improved resolution,  which is still compatible with
the pixel size chosen to represent the data.

Thus,  if  $t(\vec{x})$  is  the   total  PSF of the   observed  image
$d(\vec{x})$, and  $r(\vec{x})$  the narrower  PSF  of the deconvolved
image $f(\vec{x})$, one should apply a deconvolution kernel $s(\vec{x})$ so
that:

\begin{equation} 
t(\vec{x}) = s(\vec{x}) \ast r(\vec{x}) 
\end{equation}
where $\ast$ stands for the convolution operator.

In addition, the PSF  of the deconvolved image  is known, since it  is
the function $r(\vec{x})$, which is chosen by the  user.  We thus know
that all point sources will have the  same form $r(\vec{x})$, and this
{\em prior knowledge} may be used to write the solution as:

\begin{equation} 
f(\vec{x}) = h(\vec{x}) + \sum_{i} a_i r(\vec{x} - \vec{c_i})  
\end{equation}

where  $a_i$ and $\vec{c_i}$  are free parameters corresponding to the
intensity and position of point  source  $i$, and $h(\vec{x})$ is  the
part of   the light distribution which is   not in the  form  of point
sources.

Moreover, as  $h(\vec{x})$  itself must satisfy the  sampling theorem,
and corresponds to  an image  obtained with  a  PSF $r(\vec{x})$,  MCS
introduce a   smoothing    term which   removes the    variations  of
$h(\vec{x})$ on a scale length shorter than allowed by $r(\vec{x})$.

The MCS method has proven able to provide reliable and sometimes quite
spectacular results  (e.g., Courbin et al.\  \cite{courbin97}; Courbin
et al.\  \cite{courbin98}; Jablonka et al.\ \cite{jablonka00}; Courbin
et al.\ \cite{courbin00}).  However, it is obvious that the quality of
the deconvolution not only depends on the quality of the algorithm but
also on the accuracy of  the PSF determination,  a point which was not
thoroughly addressed in   the original MCS paper,  where  the  PSF was
basically assumed to be known.

In practice, the PSF is generally determined from  the images of point
sources (i.e.,  stars)  which are sufficiently  isolated.  However, in
some cases, the fields are so crowded that basically no isolated point
source  can be found.  The aim  of this paper  is to  present a method
which allows to determine accurate  PSFs, especially in critical cases
such as crowded fields.  Such reliable PSFs are not only essential for
carrying out meaningful deconvolution, but also for obtaining accurate
point  source photometry in  crowded  fields.  The algorithm  which is
presented here addresses both issues.

\section{Method}

In the  following, we only consider  astronomical images for which the
PSF is approximately constant throughout  the field of view.  This  is
generally not   fully true for real astronomical   images but, in most
cases, one can restrict the work to  small enough areas, where the PSF
is approximately constant.  Note that this is not a problem in crowded
stellar fields, where enough stars are available to determine the PSF,
even in relatively small subfields.

The usual way to obtain the PSF of  an astronomical image is to derive
it  from the  shape of isolated  point sources.   The total PSF 
$t(\vec{x})$ indeed
corresponds to the shape of such a point source, after recentering and
normalization to a total  intensity of 1.   This can only be done when
at least one isolated point  source of adequate intensity is available
in the  field.  There  exist two  classes  of images for   which this
simple PSF determination is not possible: (1)  when no point source is
present and (2) when there are so many point sources than none of them is
sufficiently isolated.  We shall   deal with the second  case (crowded
fields),  which is   common  in  a    large number of     astronomical
observations,   such as  the galactic   plane,  dense star clusters or
external galaxies.

Let $d(\vec{x})$ be the  observed light distribution, $f(\vec{x})$ the
light distribution  at a better  resolution  (as would  be observed at
infinite    $S/N$  with  an    instrument  of   PSF  $r(\vec{x})$) and
$n(\vec{x})$ the noise in the observed image.  Then:

\begin{equation}  
d(\vec{x}) = s(\vec{x}) \ast f(\vec{x}) + n(\vec{x}) 
\end{equation}

Let us consider a part of the  image which only contains point sources
(isolated or not).   Then, we know  that, in  the subimage considered,
the deconvolved light distribution may be written:

\begin{equation} 
f(\vec{x}) = \sum_{i} a_i r(\vec{x} - \vec{c_i})  
\end{equation}

The  classical deconvolution problem  would be: given $d(\vec{x})$ and
$s(\vec{x})$,  recover $f(\vec{x})$.  We  derive  the principle of our
PSF determination method by taking the same Eq.\ (3) and solving it in
the reverse way:  given $d(\vec{x})$ and  assuming $f(\vec{x})$ to  be
known, obtain $s(\vec{x})$.

However, while we  know that $f(\vec{x})$ may  be written in  the form
(4), we  do  not know   the  values of   the  coefficients $a_i$   and
$\vec{c_i}$.  They can thus be considered as free parameters, which will be
determined simultaneously with the PSF $s(\vec{x})$.

If we consider an image with $N$ pixels, containing $M$ point sources,
we are left with the problem of determining  $N + 3M$ parameters, i.e. 
$N$ pixel values of  the PSF and 3 parameters   for each point  source
(one intensity and two coordinates).  In practice, however, the PSF often
has significant values on an area much smaller than the field considered.
In such a case, the number of free parameters decreases significantly.

The free parameters are obtained by minimizing the following function:

\begin{equation} 
{\cal S} = \sum_{j} \frac{1}{\sigma_j^2} (d_j - [s \ast f]_j)^2 + \lambda H(s)  
\end{equation}
where $d_j$ is the intensity of the  image in pixel $j$, $\sigma_j$ is
the standard deviation in  the same pixel, $f$ is  in the form (4) and
$H(s)$  is a  smoothing  constraint which is  introduced  in  order to
regularize the  solution.  We choose  a local smoothing similar to the
one proposed in MCS:

\begin{equation} 
H(s) = \sum_{j} (s_j - [g \ast s]_j) 
\end{equation}
where $s_j$ is the value of the PSF at pixel $j$ and $g$ is a gaussian
function.   The width of $g$ and  the Lagrange parameter $\lambda$ are
adjusted so  that,  when the  function $\cal{S}$   reaches its minimum
value, the $\chi^2$ has the appropriate magnitude:

\begin{equation} 
\chi^2 = \sum_{j} \frac{1}{\sigma_j^2} (d_j - [s \ast f]_j)^2 \simeq N  
\end{equation}

As any other  inverse problem, this one is  an  ill-posed problem.  It
thus  admits  an infinity of  solutions,  and the smoothing constraint
alone does not guarantee  that the minimization of  (5) will provide a
meaningful solution.

To illustrate this, let us consider  the determination of the PSF from
the images of stars in  a crowded field and let us focus on a 
particular star in that field.   The presence of neighbouring
peaks  in the image may  be interpreted  as  due either to neighbouring
stars or to bumps in the PSF.  Even if all stars  present in the field
are considered explicitly in Eq.  (4), once  the algorithm attemps  to
minimize (5), it  could interpret part of  the light in a point source
as a bump in the  PSF of a neighbouring one.   Indeed, the function (5)
is likely to have local minima with part of the light in the center of
point sources attributed to bumps in the PSF wings.

To avoid such local minima, we proceed in the following way.

First, the PSF  is approximated  by  either an   analytic or a   known
numerical  function (e.g., a sum of  gaussians  or a Moffat function). 
These functions  are  fitted to the   point sources by   least squares
minimization.  This  first  fit  provides approximate values   for the
intensities $a_i$ and the centers $\vec{c_i}$ of the point sources, as
well as  a very  rough estimate  of the  PSF shape.  By  construction,
these analytic estimates of the PSF cannot contain bumps in the wings.

In a second step, we add a numerical component to that analytic estimate
of the PSF. In  order to avoid  the aforementioned problem of bumps in
the wings, we start by adding these numerical residuals in the central
region of the  PSF and, as the  fit proceedes, we gradually extend the
region considered.  This ensures that the
algorithm will attempt to fit the central parts  of all stellar images
by appropriate intensities in the center pixels of the PSF, and not by
bumps in the wings, since the wings are modified only after the center
intensities have been correctly fitted.  We should also mention that
the smoothing constraint (6) is applied to the numerical component
only, and not to the first analytic estimate.

In the  case of HST  images, the fit  of an  analytic  function in the
first step is replaced by the fit of  a numerical estimate of the PSF,
as computed with the Tiny Tim package (Krist and Hook \cite{TinyTim}).
So, approximate intensities and   positions are also obtained for  all
point sources  and, since the PSF  shape is  fixed  at that stage,  no
unwanted bump can  appear in the wings.  The  second step is the same as
for ground-based images: an additional numerical component is added in order 
to improve the agreement between the PSF and the observed point source images.

We should point out that the Tiny Tim software  computes the total PSF
$t(\vec{x})$ and  not $s(\vec{x})$.  In practice, we found that the results can
be improved by proceeding in two steps.  First, the kernel $s(\vec{x})$ is
determined from the synthetic image of $t(\vec{x})$ obtained with the
Tiny Tim software.  The extremely high $S/N$ of that synthetic image allows
an accurate determination of $s(\vec{x})$ even if the first approximation
(which we take as the Tiny Tim image itself) is rather far from the
solution, especially in the core.  In a second step, this approximate
$s(\vec{x})$ obtained from a deconvolution of the Tiny Tim image is 
used as a first approximation of the PSF to be determined on the
observed images, first approximation to which a numerical component will
be added.

Moreover, we should add a note on how the number of point sources, as well as
the initial guess of their positions and intensities, are determined.  We
first use a standard algorithm for detecting point source, such as the DAOFIND
task included in DAOPHOT (Stetson \cite{Stetson}).  This first guess allows to
obtain a first approximation of the PSF.  Then, the image of the residuals is
inspected, which allows to identify areas where the fit is not satisfactory. 
Additional point sources are added in these areas, until the fit becomes
satisfactory.  Note that this allows to improve the accuracy of the PSF which,
in turn, allows to detect fainter blending stars, so that the method
simultaneously converges towards higher accuracy in astrometry, photometry and
PSF recovery.

Finally, a word about the computing time.  In its present form, the program,
running on a standard PC, takes about 1 hour to completely process a $128
\times 128$ image with 25 sources and up to 10 hours with 100 sources.  It is
thus significantly slower than, e.g., DAOPHOT, which is not surprising since
we put emphasis on the accuracy of the results rather than on the speed. 
However, a considerable time is saved in the case of a photometric monitoring,
where the sources positions do not need to be recomputed each time (Gillon et
al. \cite{Gillon}).  Moreover, we are working on the algorithm for finding the
minimum of the function $\cal{S}$ (Eq.~5), which can also be significantly
improved in terms of computing time.

\begin{figure}
\centering
\includegraphics[width=8.7cm]{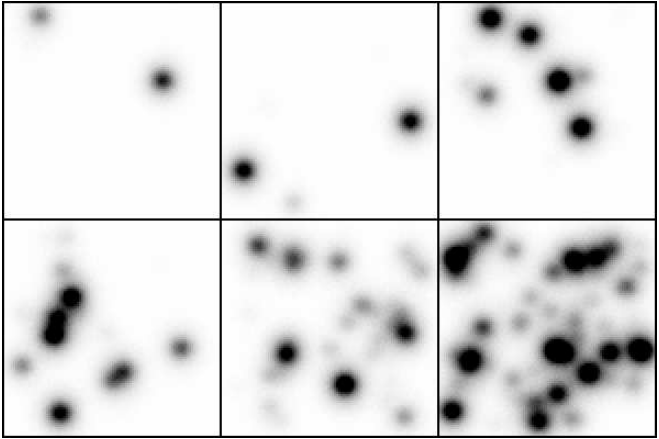}
\caption{Synthetic images used to check the PSF recovery and photometric
accuracy versus crowding.  {\it From top left to bottom right:}
Fields containing 3, 6, 12, 25, 50 and 100 stars.  Not all stars are visible on
the figure because of blending and of the 7.5 magnitude range.}
\label{fig:crowding_ima}
\end{figure}

\section{Simulated ground-based observations}

We use simulated ground-based observations in order to test the accuracy of the
PSF determination and of the photometry (1) as a function of the crowding and
(2) as a function of the $S/N$.

\subsection{Influence of crowding}

The influence of crowding is tested on six images with the same size, same
PSF and same typical $S/N$ but different numbers of stars in the field, namely
3, 6, 12, 25, 50 and 100.  All these images have $128 \times 128$ pixels, the
PSF has about 8 pixels FWHM and is constructed as the sum of two gaussians, a
first one representing the core and a second one, about two times broader,
accounting for the wings.  These two gaussians have elliptical isopohotes but
their major axes are perpendicular to each other, so that they cannot be well
fitted by a Moffat function.  This is not a very favourable case for our
algorithm, since the analytical fit provides a rather poor approximation of
the PSF and the numerical component is thus important.

The stars are positioned at random and their intensities are also randomly
generated, with a uniform distribution in magnitude and a range of 7.5 mag
(i.e. a factor 1000 in intensity).  We add a sky background of about 
30 e$^-$/pixel.  The $S/N$ in the center of the stellar images varies from less
than 1 for the faintest stars to about 70 for the brightest ones.

The six simulated images are displayed in Fig.~\ref{fig:crowding_ima}, which
shows
that we go from isolated stars up to such a crowding that all stars are more
or less severely blended: in the latter case, the average separation between
stars is close to the PSF Full Width at Half Maximum (FWHM).

\begin{figure}
\centering
\includegraphics[width=8.7cm]{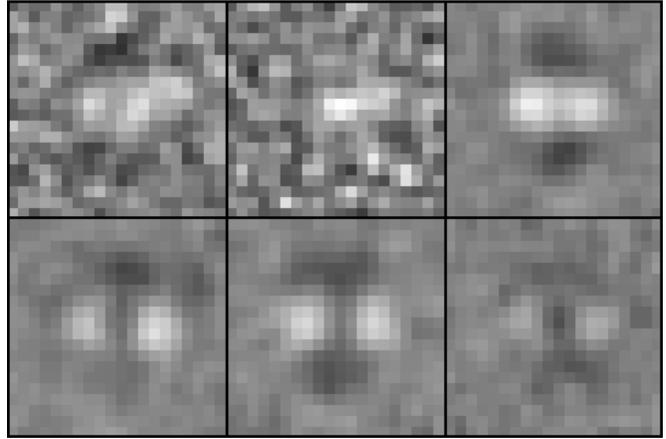}
\caption{Difference between the reconstructed PSF and the exact one, as a
function of crowding.  The grey scale goes from $-2\%$ (white) to $+2\%$
(black) of the PSF peak intensity. {\it From top left to bottom right:}  the
PSF is determined from the fields containing 3, 6, 12, 25, 50 and 100 stars.}
\label{fig:crowding_res}
\end{figure}

\begin{figure}
\centering
\includegraphics[width=8.7cm]{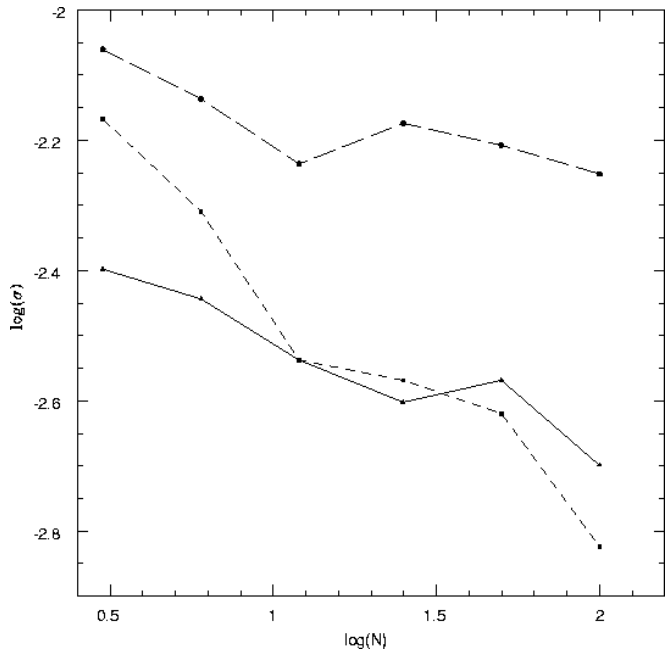}
\caption{Standard deviation of the PSF residuals (in logarithmic scale) versus
logarithm of the number of stars in the field.  {\it Long dashes:} PSF
constructed from the image of the brightest star in the field, assumed to be
perfectly centered and isolated.  {\it Short dashes:} PSF constructed from the
images of all stars in the field, assumed to be perfectly centered and
isolated.  {\it Continuous line:} Our results.}
\label{fig:crowding_plot}
\end{figure}

The differences between the reconstructed PSFs and the exact ones are shown in
Fig.~\ref{fig:crowding_res}, for the six cases considered.  The standard
deviation of these residuals is computed in a circular aperture centered on
the PSF and of 32 pixels diameter, i.e. 4 times the PSF FWHM.  This is the
area where the PSF intensities may be considered significant ($> 1\%$ of the
peak intensity).  These standard deviations, in units of the PSF peak
intensity, are plotted in Fig.~\ref{fig:crowding_plot} and compared with the
results which would be obtained in two special cases.  First, the PSF which
would be given by the image of the brightest star in the field, should it be
completely isolated and, secondly, the PSF which would be deduced from all
stars in the field, should they all be isolated.  Fig.~\ref{fig:crowding_plot}
shows, first, that the recovered PSF is always more accurate than what would
be deduced from the image of the brightest star, the improvement increasing
with the crowding of the field.  This means that the advantage provided by the
additional signal more than compensates the handicap due to blending.  In the
first four images (number of stars less than 50), our recovered PSF is even
better than the one which would be obtained by summing all the stars in the
field, if they were isolated.  Although this might appear unphysical, this is
due to the way we determine the PSF, first by the fit of an analytical
function, then by imposing a smoothing constraint on the numerical residuals. 
This results in a rather efficient correction for the photon noise, which has
Fourier components at significantly higher frequencies than our well sampled
PSF.  When the number of stars grows, the relative photon noise decreases and
its reduction by the algorithm cannot compensate anymore for the effect of
blending, which also becomes more and more severe.

The accuracy of the photometry is checked in the following way.  First, we
compute the error $\delta a_i$ in the intensity of the source $a_i$, which is
also the total flux in the stellar image since the PSF is normalized.  This
error is simply the value returned by the algorithm minus the exact value used
to build the simulated image.  Then, we compute a theoretical estimate of the
standard error by assuming that the PSF is fitted by least squares on the
image of a single (eventually blended) star, and that {\em all other stars}
have previously been perfectly fitted (i.e. with their exact positions and
intensities).  This gives:
\begin{equation} 
\frac{\sigma(a)}{a} = [ \sum_{j} (\frac{a s_j}{\sigma_j})^2 ]^{-1/2}  
\end{equation}

where the sum runs over all pixels.

Note that, in the case of a single star and no sky background, this formula
reduces to $\frac{\sigma(a)}{a} = \sqrt{a}$, i.e. pure Poisson noise, as
expected.  The extra noise due to blends is taken into account in $\sigma_j$,
in which the noise corresponding to all stars is included.  Thus, all stars
blending the star considered will contribute to an increase of the
uncertainty.  However, this formula does not take into account errors in the
photometry due to inaccurate fitting of the neighbouring stars (i.e. it
neglects the covariance terms) and is thus expected to be somewhat optimistic.

The accuracy of the photometry is quantified by:
\begin{equation} 
\chi_a^2 = \sum_i (\frac{\delta a_i}{\sigma(a_i)})^2 
\end{equation}
where the sum is over all stars in the field.  The reduced $\chi_a^2$ 
($\chi_a^2$ divided by the number of stars) is expected to be close to 1 if
everything works fine.  Indeed, it fluctuates between 0.68 and 1.62, with an
average value of 1.03, which means (1) that our photometric errors are
perfectly compatible with the noise in the image and (2) that the correlated
errors do
not contribute much to the global $\chi_a^2$.  Moreover, no significant trend
with crowding appears, meaning that the strongest crowding considered (average
separation $\sim$ FWHM of the PSF) is not severe enough to jeopardize our
photometric results.

\begin{figure}
\centering
\includegraphics[width=8.7cm]{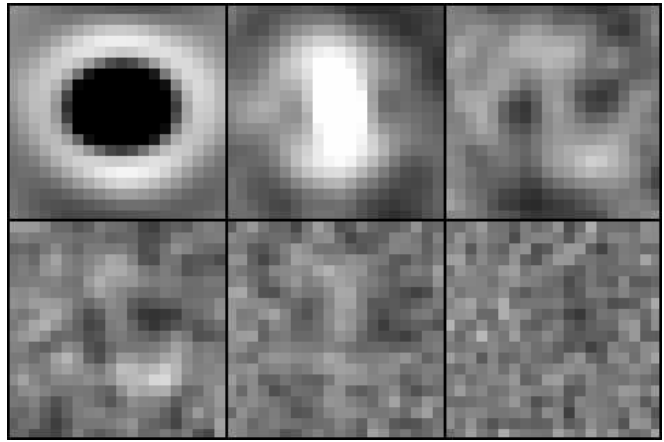}
\caption{Difference between the reconstructed PSF and the exact one, as a
function of $S/N$.  The grey scale goes from $-2\%$ (white) to $+2\%$ (black)
of the PSF peak intensity. {\it From top left to bottom right:}  the PSF is
determined from fields with increasing $S/N$ (see text).}
\label{fig:SN_res}
\end{figure}

\subsection{Influence of $S/N$}

The influence of $S/N$ on the PSF accuracy is tested in a similar way.  We use
the image with 50 stars discussed in the previous subsection and vary the
exposure level over a large range, i.e. from 30 to 10000 e$^-$/pixel in
the center of the brightest star.  This corresponds to an integrated $S/N$
varying from about 20 to 1000 for the brightest star in the field.

The differences between the reconstructed PSFs and the exact ones are shown in
Fig.~\ref{fig:SN_res}.  The standard deviation of
these residuals is computed in the same circular aperture as above and
compared with the same 2 cases as in the previous subsection.  
Figure~\ref{fig:SN_res} shows that our results are always better than the ones
which would be obtained from the brightest star, in case it would be isolated. 
At low to moderate $S/N$ (brightest star peak intensity $< 1000$ e$^-$,
integrated $S/N < 300$), our results are also better than what would be
obtained from summing all stellar images, assuming they were isolated.  Again,
this is due to the fact that the smoothing term reduces the effect of the
photon noise.  At high $S/N$ (400 to 1000), the contribution of the photon
noise becomes relatively smaller and its reduction does not compensate anymore
for the degradation due to crowding.  Nevertheless, the accuracy of the
recovered PSF continuously improves with increasing $S/N$, as expected.

The accuracy of the photometry as a function of $S/N$ is quantified in the same
way as in the previous subsection.  As the $S/N$ increases, the photometry gets
more and more accurate, but not quite as much as Eq.~(8) predicts.  Indeed,
this equation does not take into account errors coming from inaccuracies in
neighbouring stars (correlated errors).  At low $S/N$, these correlated errors
are much lower than the random ones.  However, as the random errors obviously
decrease with increasing $S/N$, the correlated errors start to play a role, and
the reduced $\chi_a^2$ grows from $\sim 1$ at low $S/N$ to 1.5 at $S/N \sim 500
$ and 1.9 at $S/N \sim 1000$ (the indicated $S/N$ are integrated ones for the
brightest star in the field).  Thus, even at the highest $S/N$, the
photometric errors are only about $40\%$ larger than would be
expected in the ideal case (zero correlation).  We should also note that they
can be reduced if the stars'positions can be constrained from many
observations, as is the case in a photometric monitoring.

\begin{figure}
\centering
\includegraphics[width=8.7cm]{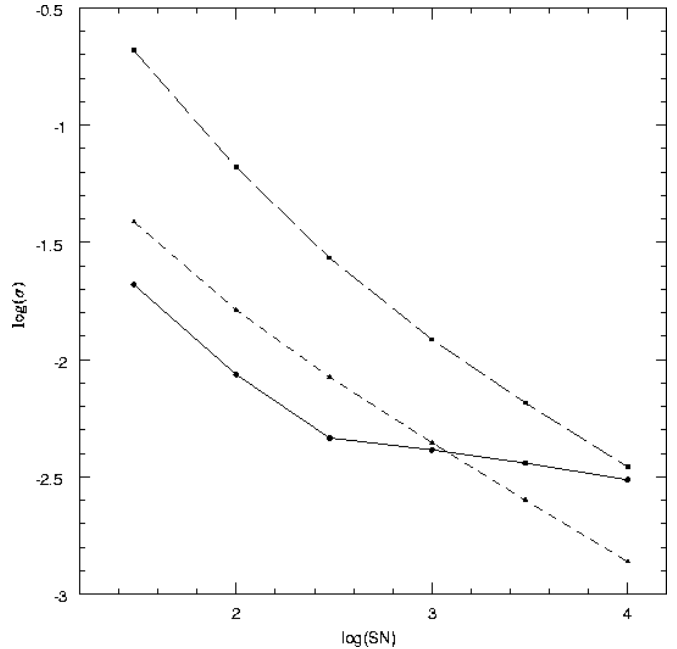}
\caption{Standard deviation of the PSF residuals (in logarithmic scale) versus
logarithm of peak intensity (see text).  {\it Long dashes:} PSF constructed
from the image of the brightest star in the field, assumed to be perfectly
centered and isolated.  {\it Short dashes:} PSF constructed from the images of
all stars in the field, assumed to be perfectly centered and isolated.  {\it
Continuous line:} Our results.}
\label{fig:SN_plot}
\end{figure}

\begin{figure}
\centering
\includegraphics[width=8.7cm]{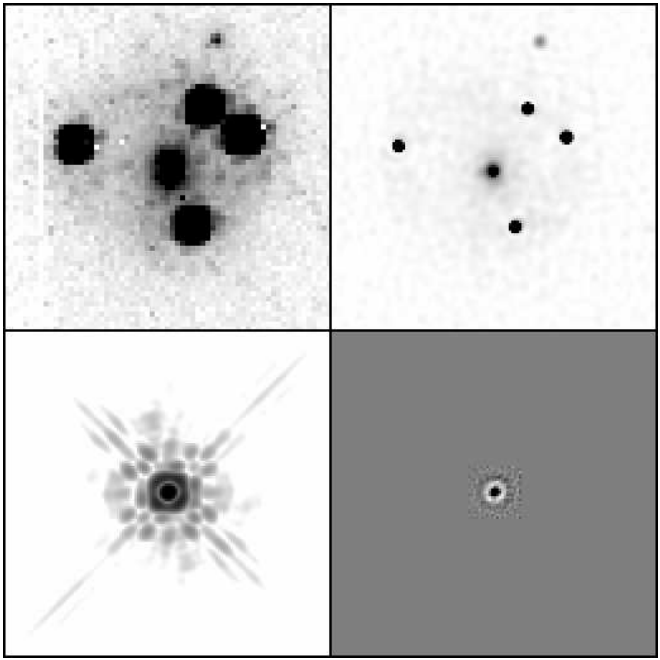}
\caption{{\it Top left:} H-band (F160W) HST/NICMOS observation of the
gravitational lens WFI2033$-$4723. {\it Top right: } simultaneous
deconvolution of 4 such images, on a $2 \times 2$ smaller pixel grid with 3
(small) pixels FWHM, shown on the same intensity scale.  {\it Bottom left: }
PSF kernel $\vec{s(x)}$ determined from a deconvolution of the Tiny Tim PSF. 
A logarithmic intensity scale is used to enhance the faint structures.  
{\it Bottom right: } numerical component added to PSF, as determined by the
present method.  In this panel, the zero level is grey, lighter areas
correspond to negative differences and darker ones to positive ones, the grey
scale goes from $-10\%$ to $+10\%$ of the PSF peak intensity.}
\label{fig:WFI}
\end{figure}



\section{PSF determination and simultaneous deconvolution of several images}

One  of the main  applications of the  MCS algorithm, in  its original
form, was to deconvolve not only an image  of a given object, but also
a  whole set  of  images of  the  same object.  In   such  a case  the
deconvolution process computes  a {\it  unique} sharpened  image  {\it
  compatible with all the images of the dataset simultaneously}, hence
the name {\it   simultaneous} deconvolution.  This  specificity of the
MCS algorithm has been transposed to the present  method. It allows to
determine {\it simultaneously} the PSFs for a series of images.  If the
individual images have been taken at different epochs, the intensities
of all point  sources in  the field  are left free  during the
deconvolution process, leading to the construction  of a genuine light
curve for all objects.

In such a case, instead of seeking the minimum of a function 
${\cal S}$ which is the sum of a $\chi^2$ and a smoothing term (Eq.~5),
one seeks the minimum of a function which is the sum of a $\chi^2$
for each of the images considered plus one smoothing term per image.
The PSFs and the source intensities are allowed to vary from image
to image, but the source positions are common to all images.  A
translation of the whole image is however allowed, to account for
different centerings of the various exposures.  All these parameters
(source positions, source intensities, translations and PSF shapes)
are simultaneously determined by the minimization algorithm.

\subsection{HST/NICMOS observations of a gravitational lens}

Even if the method presented here assumes that  the data only contain
point sources, one can adapt the algorithm to allow it to take into
account faint diffuse objects.

WFI2033$-$4723, a quadruply imaged quasar lensed by a distant galaxy,
provides a typical case where such a strategy can be  applied. 
The top left panel of Fig.~\ref{fig:WFI} shows a $H-$band HST/NICMOS 
observation of that gravitational lens.

A deconvolution of the Tiny Tim PSF (Krist and Hook \cite{TinyTim}) is used as
a first approximation of $s(\vec{x})$, which is then modified to provide the
best possible  deconvolution of the input image, which is first assumed to
contain only four point sources.  As the image actually contains a diffuse
background (the lensing galaxy) in addition to the point sources, part of that
background is interpreted as increased PSF wings and another part appears in
the residuals (observed image minus reconvolved model).  This latter part may
be subtracted from the original image, which thus allows to correct for a part
of the diffuse background.  That corrected image may then be used to obtain
an improved PSF.

Applying that procedure iteratively allows to obtain a PSF which, at each
iteration, contains a smaller contamination by the diffuse background.
Once that procedure has converged, the PSF is used as input for deconvolution
of the original image by the classical MCS algorithm, including a diffuse
background.

The result is shown in the top righ panel of Fig.~\ref{fig:WFI}.  The bottom
left panel of Fig.~\ref{fig:WFI} shows the PSF kernel $s(\vec{x})$
obtained from deconvolving the PSF computed with the Tiny Tim software while
the bottom right panel shows the numercial component which has to be added to
the former PSF kernel in order to obtain a good deconvolution of the input
image.

This method has been applied to more complex cases, as the Cloverleaf
gravitational lens H1413+117 (Magain et al. \cite{Magain88}).  In that case,
it allows to detect the lensing galaxy as well as part of an Einstein ring,
which is completely hidded in the original data (Chantry, Magain \& Courbin,
in preparation).

\subsection{Example: detailed light curve of a faint Mira in the halo of
  NGC5128, Cen A}

With the increasing performances of  modern telescopes, it becomes
possible to study the stellar populations of  objects that were so far
unresolved under standard seeing conditions,  around 1\arcsec. One can
even  start to resolve (nearby)  galaxies into stars  and to construct
actual colour-magnitude diagrams.  However, while the resolution of the
observations improves, the field   of view often decreases, making  it
very difficult  or even impossible  to  observe at the same  time  the
field  of interest and  the relatively bright, isolated stars 
generally required to build the PSF.

The search for faint blended variable stars in  nearby galaxies is one
of the topics  directly influenced by  the  quality of the available PSF. 
Rejkuba et al. (\cite{Marina}) have found  dozens of Mira stars in the
halo of  the giant elliptical  galaxy NGC~5128 (Centaurus A), but also
note that strong  blends  often  hamper  the accurate  measurement  of
periods.  This  occurs  because the Mira  itself can   be in  a close
blend, but also often because the field where it lies is far away from
any suitable PSF star. Our method computes the PSF directly from all the
stars, blended or not, in the field of view, hence taking advantage of
the total $S/N$ of all available point  sources.  The PSF is
also well  representative of the instrumental/atmospheric  blurring at
the position of the object of interest in the image.

\begin{figure}
\centering
\includegraphics[width=8.7cm]{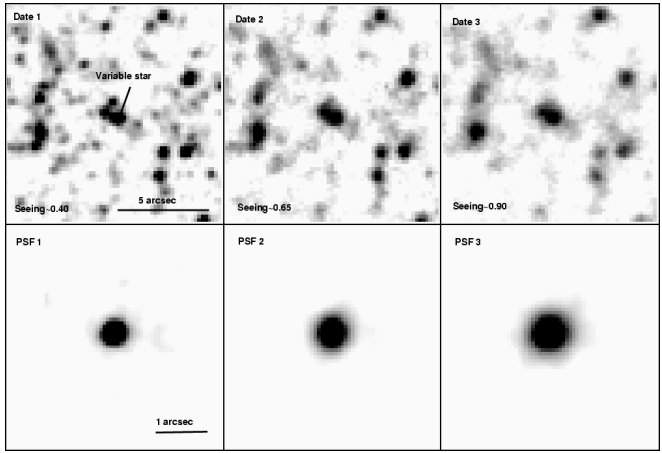}
\caption{Zoom on a Mira star in the halo of the radio galaxy NGC 5128
  (Centaurus A),  selected from  Rejkuba  et al.  \cite{Marina}. Three
  VLT K-band images with very different seeing conditions are shown  
  on the top panels.  The exposure time  is the same in
  all three exposures: 3600sec.  The pixel  size is 0.144\arcsec.  The
  three PSFs in the bottom panels were obtained using the simultaneous
  deconvolution algorithm   presented here.  The   pixel size  is now
  0.072\arcsec.  These  PSFs  are  the PSF kernels  to be  used  in  the  MCS
  algorithm to improve the  data from their  original resolution, to a
  better and  {\it   common}  resolution  of  3   pixels   FWHM, i.e.,
  0.22\arcsec.}
\label{fig:Mira_field}
\end{figure}

\begin{figure}
\centering
\includegraphics[width=8.7cm]{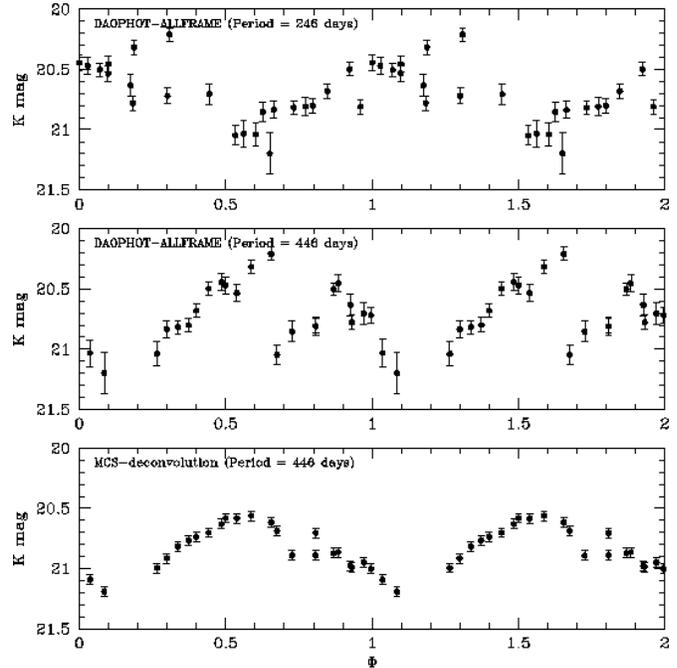}
\caption{{\it Top panel:} Light curve (shown here as phase diagram) 
  of the Mira presented in Fig.~\ref{fig:Mira_field}, as obtained with
  standard PSF  fitting  techniques (DAOPHOT-ALLFRAME).   The PSFs  are
  computed   independently in  the  individual  frames from relatively
  isolated stars in the field of view, well  outside the field
  of Fig.~\ref{fig:Mira_field}.    The 1$\sigma$   error bar    on the
  photometric points is $\sim 0.15$ mag.  The period found from Fourier
  techniques  is  246 days.  {\it Middle panel:} Phase diagram constructed 
  from the same data points, but with the 446 days period determined from our
  improved photometry.    {\it   Bottom panel:}  Phase
  diagram obtained from the same images and the algorithm presented here.
  The 1$\sigma$  error  bar is   now 0.05  magnitude and  the  scatter
  between the points is drastically improved. The new period
  measured is 446 days, completely different from the period measured
  with standard photometric techniques.}
\label{fig:Mira_curve}
\end{figure}

We have selected one of the Mira candidates found  by Rejkuba et al.   
(\cite{Marina})  using VLT  K-band images.   The  region used  for the
deconvolution      is a  subset   of    the    whole image obtained.   
Fig.~\ref{fig:Mira_field} presents  a  sample of  three  images of the
same field,  taken at different  epochs, with very different  seeings. 
The photometry of the  Mira star has been  carried out on the  21 data
frames   available   and a phase   diagram   could  be constructed, as
displayed in the  bottom panel of Fig.~\ref{fig:Mira_curve}.  The photometric
uncertainties, estimated from the scatter in the light curve of a constant
star of the same magnitude as the Mira, amount to 0.05 mag, as compared to an
average of 0.15 mag with the PSF-fitting method.  Not only
the error bars on the individual  photometric   points improve  by a
factor of almost 3 when compared with classical methods (top and middle  panels
of Fig.~\ref{fig:Mira_curve}),  but  also   the  period  obtained is
different.  The frequency peaks  found in the  Fourier analysis of the
light  curves with the present algorithm are much  stronger than with the
classical method. The  446 day period  is  clearly pointed  out by the
Fourier analysis, while the 246  day period supported by the classical
PSF fitting no more yields a prominent peak.

\section{Conclusion}

The deconvolution-based algorithm presented here has   been  designed to
compute
accurate PSFs in fields very crowded with point sources.  It not only
computes the PSF but also provides the  photometry and astrometry of
all point sources in the field of view.   The algorithm can be applied
to  single images or to  a set of  images of a   given field.  In the
latter case, the images are processed  simultaneously, in the sense of
the MCS deconvolution  algorithm: a PSF is computed  for each image,
by considering all images simultaneously.
The photometry of  all points sources  is obtained for all
images in the dataset, i.e., light curves are directly produced.

The method is clearly useful when few or no isolated PSF stars
are available  in the field of  view, in the  case of extreme crowding
and in the case of strong PSF variation  accross the field (in which case the
PSF has to be determined from stars very close to the target). It is also
very efficient in extracting  photometric information from datasets 
of very   heterogeneous quality (e.g., with  a  broad range  of seeing
values). In this case, the astrometry of the points sources in the best
seeing data effectively constrains the astrometry  of the bad seeing data, as
long as all the data are processed simultaneously. 

Although the present method  is  not designed to   handle images that consist
in a
mixture of  point sources  and extended ones,  it can cope  with extended
objects  which  are faint in comparison to the  point sources, by
running the algorithm in an iterative way. 

\begin{acknowledgements}
The authors would like to thank  Marina  Rejkuba for providing the VLT images
of the Mira stars  presented  in  Fig.~8,  as  well as for  all  the  
technical  details  on  how  they were  reduced   and calibrated. This work
has been supported by ESA and the Belgian Federal Science Policy Office under
contracts PRODEX 90195 and P\^ole d'Attraction Interuniversitaire P5/36.  FC
acknowledges financial support from the Swiss National Science Foundation
(SNSS).
\end{acknowledgements}

\end{document}